\documentclass[letterpaper, 10 pt, conference]{ieeeconf}

\IEEEoverridecommandlockouts
\usepackage{amsmath,amssymb,amsfonts}
\usepackage{array}
\usepackage{booktabs}
\usepackage{hyperref}
\usepackage{cite}
\let\labelindent\relax
\usepackage{algorithm}
\usepackage{algpseudocode}
\usepackage{enumitem}
\usepackage{comment}
\usepackage{graphicx}
\usepackage{xcolor}

\DeclareMathOperator*{\argmin}{arg\,min}

\title{\LARGE \bf Holding the Cold: Retention-Aware Inventory Control for Industrial Refrigeration}

\author{Lily Y. Chen, Vade Shah, Dan Walsh, and Jason R. Marden
\thanks{This work is supported by the California Energy Commission (project \#EPIC-24-012) and the NSF GRFP (grant \#2139319).}
\thanks{L. Y. Chen (\texttt{yuxuanchen@ucsb.edu}), V. Shah, and J. R. Marden are with the Department of Electrical and Computer Engineering at the University of California, Santa Barbara, CA. Dan Walsh (\texttt{danthewalsh@gmail.com}) is with CrossnoKaye\textregistered.}
}

\begin{document}
\maketitle
\thispagestyle{empty}
\pagestyle{empty}

\begin{abstract}
Industrial refrigeration consumes substantial electricity, and compressors are the dominant energy consumers in these systems, making their operation central to improving energy efficiency. Compressors are most efficient when operated at full capacity, but the cooling they provide at full capacity can exceed the facility’s immediate needs. This motivates pre-cooling, where compressors operate at full capacity and the excess cooling is stored as thermal inventory for future heat loads. However, stored cooling is inherently lossy: colder spaces attract additional heat from their surroundings, creating a tradeoff between maximizing compressor efficiency and avoiding cooling that dissipates before it can be used. We study this tradeoff through a retention factor, defined as the fraction of thermal inventory that survives after one minute, with the goals of understanding how retention affects the value of pre-cooling and designing effective control policies that account for this lossiness. We formulate compressor control as a stochastic inventory problem, solve for an optimal policy via dynamic programming alongside two simpler alternatives, and evaluate the performance of these policies using models fit from real industrial refrigeration facility data. Our results show that (i) pre-cooling can provide substantial energy savings when thermal inventory is sufficiently persistent, but that failing to account for retention can make aggressive pre-cooling increasingly costly as losses grow, and (ii) structurally simple policies that account for lossiness can still achieve near-optimal performance in lossy systems.
\end{abstract}

\section{Introduction}

Cooling and refrigeration consume approximately 17\% of global electricity \cite{dong2021}, a metric that is expected to increase with the rapid expansion and growth of data centers \cite{iea2018cooling}. The sector is therefore a prime target for new energy-saving technologies, including not only improvements in physical infrastructure, but also better decisions about how to operate a refrigeration system \cite{jain2011,koeln2014,larsen2006}.

\begin{figure}
    \includegraphics[width=\columnwidth]{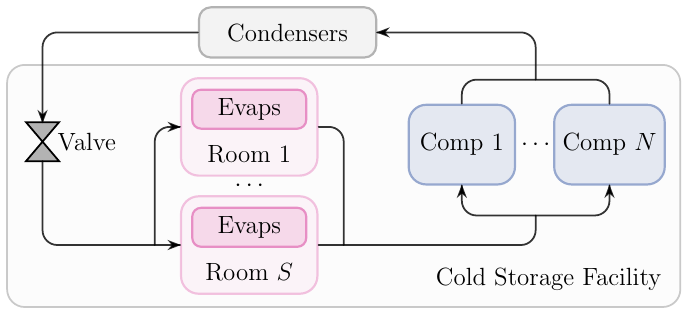}
    \caption{Simplified schematic of an industrial refrigeration facility. A refrigerant travels in the direction specified by the arrows; evaporators transfer heat from the rooms to the refrigerant (pink). It then travels to the compressors (blue), which pressurize it and enable heat transfer to the external environment via the condensers.}
    \label{fig:facility}
\end{figure}

Making better cooling decisions requires understanding the basic operating principles of the refrigeration cycle. Depicted in Figure~\ref{fig:facility}, the standard vapor-compression cycle circulates a refrigerant through a series of mechanical devices. The refrigerant is the medium through which heat is transferred from a cooled space to the external environment. At the heart of this cycle is the compressor, the main device that thermodynamically enables the heat transfer to the environment by pressurizing the refrigerant. Doing so takes significant energy: compressors account for about 70\% of the energy used in industrial refrigeration systems \cite{lekov2009}.

For energy savings, the key question becomes: how should one control the compressors so as to meet cooling demands while minimizing energy consumption? 
Much research has sought to answer this question using model predictive control \cite{hovgaard2013,afram2014,yao2021}, inventory control \cite{shah2026,john2025}, and data-driven approaches \cite{vishwanath2019}.
Underlying most refrigeration control strategies is the indispensable insight that running a compressor at full capacity is most efficient, meaning that it minimizes the ratio of work to heat removal \cite{stoecker1998}.

One might therefore be inclined to simply operate the compressors at full capacity whenever cooling is needed; refrigeration systems typically have an upper limit on temperature, but many have no lower limit. Thus, if the space is getting too warm, one can run the compressors at full capacity, and any excess cooling can be stored in the physical goods in the space \cite{altwies2002} in anticipation of future heat loads \cite{braun2003,sun2013}. In this manner, whenever a compressor is operated, it is operated at its peak efficiency, so average work over the long run should be minimized.

Unfortunately, this approach does not take into account the fact that stored cooling is inevitably lossy---heat naturally flows from warmer to colder spaces. 
Furthermore, the colder a space is kept, the larger the temperature difference with its surroundings and hence the greater the heat gain into the refrigerated space \cite{incropera2017}. These losses are hard to avoid: once started, a compressor must typically stay on for a minimum on-time \cite{acerbi2020}, so full-capacity operation naturally produces more cooling than is immediately needed.

The ability of a space to retain cooling is hence central to the tradeoff between maximizing compressor efficiency and avoiding superfluous cooling. In this work, this ability is captured by the \emph{retention factor}: the fraction of stored cooling that remains after one minute. When the retention factor is one, pre-cooling at full load is effective; however, in practice, this factor is strictly less than one, so full load operation may not always be desirable. Understanding the relationship between the retention factor and energy requirements crucially informs control strategies and infrastructure investments, warranting a deeper study on its effects.

\subsection{Contributions}

This paper focuses on (1) understanding the value of pre-cooling as the retention factor changes and (2) designing simple, effective control policies that account for the retention factor. To formally study the interplay between pre-cooling and the retention factor, we first develop a model of the stored cooling itself, which we call the \emph{thermal inventory}. 
The thermal inventory behaves much like the stock in inventory control \cite{shah2026,john2025}, a cornerstone problem in operations research \cite{arrow1951,scarf1960}: it is replenished by the compressors and depleted by random heat loads arriving at the facility.
Importantly, only a fraction of the inventory, given by the retention factor, carries over from one time step to the next. The resulting optimization problem is then that of minimizing the work of the compressors subject to the constraint that the thermal inventory remains nonnegative.

We then propose three policies as solutions to this problem. First, we formulate a dynamic program that minimizes expected compressor energy, and solve it over discretized states and actions to obtain an approximately optimal control policy. Because this policy can be hard to interpret and expensive to compute, we also develop two simple policies as alternatives to the dynamic program: taking advantage of efficiency at full capacity, the full-load (FL) policy switches compressors between off and full capacity; inspired by the structure of the dynamic programming solution, the partial-load (PL) policy also runs them at partial load to avoid excess pre-cooling.

Finally, we apply our model and policies in simulation. Using data from a real industrial refrigeration facility in the United States, we fit models of the compressors and of the distribution of incoming heat loads at the facility. We use these models to evaluate the performance of the three policies and to understand the effect of the retention factor. The key qualitative insights are that (1) the retention factor has significant impact on the energy required to keep the facility cool, (2) full-load operation is effective when the retention factor is near one but quickly worsens as the retention factor decreases, and (3) structurally simple policies that account for lossiness can still achieve near-optimal performance in lossy systems.

\section{Problem Setting}
\label{sec:problem_setting}

Consider a cold storage facility with a set of rooms denoted by $\mathcal{R} = \{r_1, \dots, r_S\}$. At each one-minute time step $k \in \{0, \dots, N - 1\}$, one seeks to operate the refrigeration system so as to keep the temperature in each room below a prespecified temperature limit $T^{\max}$. Illustrated in Figure~\ref{fig:facility}, the basic operation of the refrigeration cycle can be described as follows:

\begin{itemize}
    \item In each room $r$, a set of evaporators $\mathcal{E}_r$ transfer heat from the room to a refrigerant, causing it to boil;
    \item A central set of compressors $\mathcal{C} = \{c_1, \dots, c_M\}$ pressurizes the refrigerant vapor collected from all of the rooms;
    \item The condensers reject heat from the refrigerant to the outside;
    \item The expansion valve depressurizes the refrigerant.
\end{itemize}

Within this cycle, the compressors are the key enablers of heat transfer: raising the refrigerant's pressure (and temperature) is what allows the condensers to naturally transfer heat to the external environment. As such, they are also the dominant energy consumers in the cycle, so the focus of this work is on operating them efficiently while still respecting the temperature limits in each room.

\subsection{Compressor behavior}

We begin by developing our mathematical model of a compressor. Efficient compressor operation depends on two ingredients. The first is the compressor \emph{work--heat} curve $W_c(q_{c,k})$, which maps heat removal to electrical work (in kWh). At each time $k$, each compressor $c \in \mathcal{C}$ can remove
\begin{equation}
    q_{c, k} \in \mathcal{Q}_c \triangleq \{0\}\cup[q_c^{\min},q_c^{\max}]
    \label{eq:v3-compressor-range}
\end{equation}
kilowatt-hours (kWh) of heat from the refrigerant stream. Here, $q_{c, k} = 0$ indicates that the compressor is off, and the positive quantities $q_c^{\min}$ and $q_c^{\max}$ are the minimum and maximum amount of heat the compressor can remove when turned on, respectively. The amount of work required to remove $q \in \mathcal{Q}_c$ kWh of heat is then well-modeled by
\begin{equation}
    W_c(q)
    =
    \begin{cases}
        0 & q = 0 \\
        W_c^{\min}
        +
        \frac{\Delta W}{\Delta q}
        \bigl(q-q_c^{\min}\bigr), & q \in [q_c^{\min},q_c^{\max}].
    \end{cases}
\end{equation}
Here, $\Delta q = q_c^{\max}-q_c^{\min}$ and $\Delta W = W_c^{\max}-W_c^{\min}$, where $W_c^{\min}$ and $W_c^{\max}$ are the amount of work the compressor consumes at minimum and maximum capacity, respectively. We assume $W_c^{\max} / q_c^{\max} < W_c^{\min} / q_c^{\min}$, meaning that the compressor is most efficient when run at maximum capacity \cite{stoecker1998}. A work-heat curve for a real compressor is shown in Figure~\ref{fig:everett-data-fit}.

The second ingredient is a minimum on-time: to prevent damage from repeated switching \cite{acerbi2020}, whenever a compressor is turned on, it must remain on for at least $H_c > 0$ consecutive minutes. Formally, let
\begin{equation}
    y_{c,k}(q_{c,k})
    =
    \begin{cases}
        1 & q_{c,k}>0,\\
        0 & \text{otherwise}
    \end{cases}
    \label{eq:v3-compressor-status}
\end{equation}
denote on/off status (written $y_{c,k}$ when the dependence on $q_{c,k}$ is clear), and let $\tau_{c,k}\in\{-1,0,\dots,H_c-1\}$ denote the remaining on-time, with $\tau_{c,k}=-1$ meaning the compressor is off. Then, the minimum on-time is enforced by
\begin{equation}
    y_{c,k} \geq \min \{1, \tau_{c, k} \},
    \label{eq:v3-hold-constraint}
\end{equation}
while $\tau_{c,k + 1}$ itself evolves according to the dynamics
\begin{align}
    \tau_{c,k + 1} &= h_c(\tau_{c, k}, q_{c, k}) \nonumber \\
    &\triangleq
    \begin{cases}
        \tau_{c,k}-1
            & \tau_{c,k}>0,\\[2pt]
        \bigl((1-H_c)\tau_{c,k}+1\bigr)y_{c,k}-1
            & \tau_{c,k}\le 0.
    \end{cases}
    \label{eq:v3-timer-dynamics}
\end{align}

Taken together, these two ingredients suggest a simple strategy---turn the compressor on at $q_c^{\max}$ and leave it there for the entire minimum on-time. Doing so typically removes more heat than the facility needs over that window and \emph{pre-cools} the rooms below $T^{\max}$; this cooling buffer then allows one to turn the compressor off altogether for a period of time after the minimum on-time has passed.

An issue with this strategy is that pre-cooling is not free. Holding the rooms colder than $T^{\max}$ increases the temperature difference driving heat into the facility, so some of the stored cooling dissipates before it can be used. Hence, an effective control strategy must balance between maximizing compressor efficiency and minimizing superfluous cooling.

\subsection{Thermal inventory}
\label{sec:inventory}

The focus of this paper is on understanding the interaction between pre-cooling and lossiness in industrial refrigeration. This involves answering the following questions:
\begin{enumerate}[label=(\roman*)]
    \item Should we perform pre-cooling?
    \item If so, when should we pre-cool, and by how much? 
\end{enumerate}
Addressing these requires a model of both pre-cooling and lossiness. We develop these through the lens of inventory control \cite{shah2026,john2025}, where the inventory level represents the \emph{aggregate} stored cooling (in kWh) in excess of the amount necessary to keep all of the rooms at $T^{\max}$.
%
%
We take $x_k = 0$ to mean that every room sits at its temperature limit, so we require
\begin{equation}
    x_k \geq 0,
    \quad
    k \in \{0,\ldots,N\}
    \label{eq:v3-inventory-feasibility}
\end{equation}
which captures the requirement that every room remain at or below its temperature limit.
Letting $q_k = (q_{1, k}, \dots, q_{M, k})$, we can then model the dynamics of this thermal inventory as
\begin{equation}
    x_{k+1}
    =
    g(x_k, q_k, q_{\mathrm{in}, k})
    \triangleq
    \rho x_k
    +
    \sum_{c\in\mathcal{C}} q_{c,k}
    -
    q_{\mathrm{in},k}.
    \label{eq:v3-inventory-dynamics}
\end{equation}
Here, \(q_{\mathrm{in},k}\sim Q_{\mathrm{in}}\) is the random, exogenous system-level heat load over the interval, measured in kWh, where \(Q_{\mathrm{in}}\) is supported on \([0,q_{\mathrm{in}}^{\max}]\), and the heat loads are taken to be independent across \(k\). The \emph{retention factor} $\rho\in[0,1]$ is the fraction of inventory retained over one minute: as $\rho$ decreases, more of the stored cooling is lost to the environment. This proportional loss is analogous to deterioration in perishable-inventory models \cite{nahmias1982}. Importantly, these are two distinct depletion mechanisms: $q_{\mathrm{in},k}$ is the external load at the temperature limit $T^{\max}$, while $(1-\rho)x_k$ is the additional load induced by carrying thermal inventory. 

Combining the operational constraints, dynamics, and inventory requirements, we can state the compressor cooling problem studied in this work as follows:
\begin{align}
    \min_{\{q_{c,k} \in \mathcal{Q}_c \}}
    \quad
    &
    \sum_{k=0}^{N-1}
    \sum_{c\in\mathcal{C}} 
    \, \mathbb{E}
    \left[
        W_c(q_{c,k})
    \right]
    \label{eq:v3-energy-optimization}
    \\
    \text{s.t.}
    \quad
    &
    \eqref{eq:v3-compressor-status}\text{--}\eqref{eq:v3-inventory-dynamics}.
    \nonumber
\end{align}
Solutions to this problem describe whether and how one should operate the compressors to facilitate pre-cooling. In the following section, we discuss these solutions in the form of policies.

\section{Algorithms}
\label{sec:algorithms}

In this section, we present various \emph{policies}---feedback maps from the system state to compressor cooling commands---as solutions to \eqref{eq:v3-energy-optimization}. For notational convenience, let $\tau_k = (\tau_{c_1, k}, \dots, \tau_{c_M, k})$, and let
\begin{equation}
    (x_{k+1},\tau_{k+1})=f(x_k,\tau_k,q_k,q_{\mathrm{in}, k})
    \label{eq:v3-joint-dynamics}
\end{equation}
denote the combination of the dynamics in \eqref{eq:v3-timer-dynamics} and \eqref{eq:v3-inventory-dynamics} over the joint state $(x_k, \tau_k)$. A policy is a sequence of mappings $\pi=(\pi_0,\dots,\pi_{N-1})$ from the joint state to a set of compressor cooling commands:
\begin{equation}
    q_k = \pi_k(x_k, \tau_k) \in \mathcal{Q}_k(\tau_k),
    \qquad
    k \in \{0,\dots,N-1\},
\end{equation}
where
\begin{equation}
    \mathcal{Q}_k(\tau_k) \triangleq \Bigl\{ q \in \textstyle\prod_{c} \mathcal{Q}_c : q \text{ satisfies \eqref{eq:v3-hold-constraint}} \Bigr\}
    \label{eq:admissible-actions}
\end{equation}
is the set of admissible cooling commands. Below we present three such policies: an approximately optimal policy obtained from a dynamic program, and two simpler policies built around full-load efficiency, one of which also draws on the structure of the dynamic programming solution.

\subsection{Dynamic Program}

\begin{figure*}
    \includegraphics[width=\textwidth]{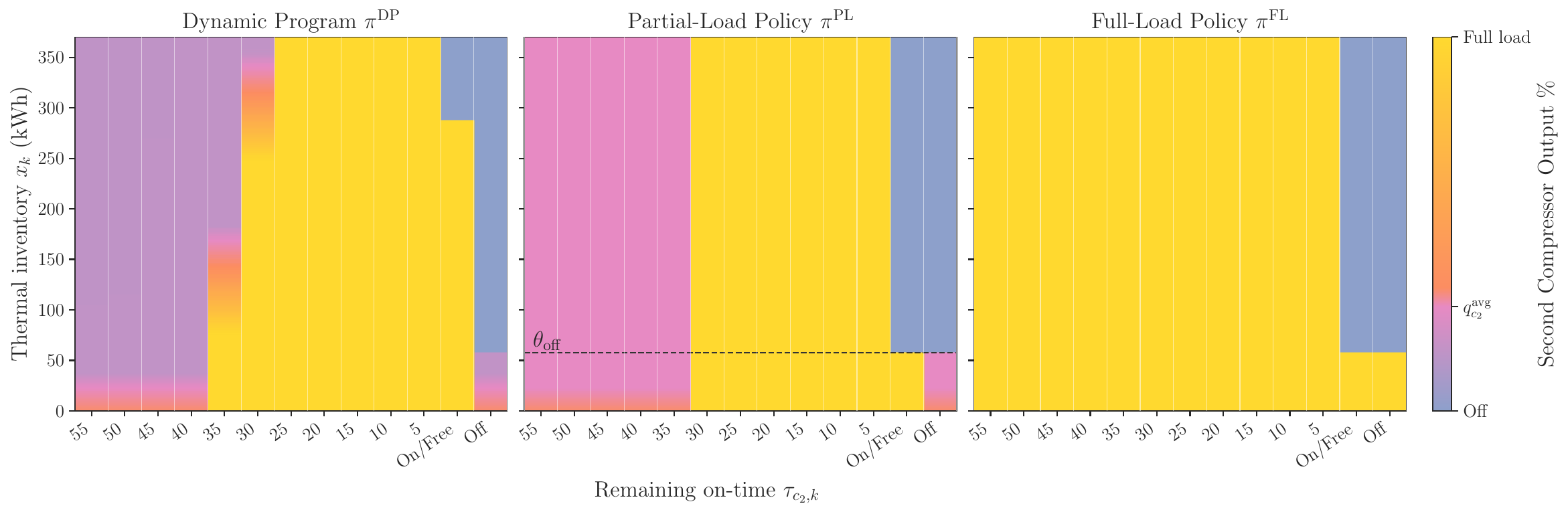}
    \caption{Visualizations of the policies $\pi^{\rm DP}$ (left), $\pi^{\mathrm{PL}}$ (center), and $\pi^{\mathrm{FL}}$ (right) for the second compressor $c_2$, whose minimum on-time is $H_{c_2}=60$ minutes; the policies are derived using real data from a cold storage facility (see Section~\ref{sec:case_study}). On each plot, the horizontal axis corresponds to the compressor's remaining on-time $\tau_{c_2, k}$, the vertical axis corresponds to the inventory level $x_k$, and the color indicates the cooling command for the second compressor $q_{c_2, k}$ as a percentage of its maximum capacity $q_{c_2}^{\max}$. Note that all three policies operate at maximum capacity when little on-time remains, but only $\pi^{\rm DP}$ and $\pi^{\mathrm{PL}}$ operate at partial capacity when much of the on-time remains.}
    \label{fig:policy-plots}
\end{figure*}

The first approach we consider is an optimal policy $\pi^*$ obtained as the solution to a dynamic program. Since the objective function \eqref{eq:v3-energy-optimization} is separable over each time step $k$, one can compute an optimal policy by solving for the optimal control action at each state and time step:
\begin{align}
    J_N(x_N, \tau_N) &= \begin{cases}
        0 & x_N \geq 0 \\
        \infty & x_N < 0,
    \end{cases} \label{eq:dp-eq-1} \\
    J_k(x_k, \tau_k) &= \begin{cases}
        \displaystyle\min_{q \in \mathcal{Q}_k(\tau_k)}
        \, G_k(x_k, \tau_k, q)
        & x_k \geq 0 \\
        \infty & x_k < 0,
    \end{cases} \label{eq:dp-eq-2} \\
    G_k(x_k, \tau_k, q) &= \sum_{c \in \mathcal{C}}
    W_c(q_c) + \mathbb{E}\!\left[
         J_{k + 1}\bigl(f(x_k, \tau_k, q, q_{\mathrm{in}, k})\bigr)
    \right]. \label{eq:dp-eq-3}
\end{align}
Here, $J_k(x_k, \tau_k)$ represents the expected cost-to-go from the state $(x_k, \tau_k)$. These equations are solved via backward recursion; starting from stage $N - 1$, one identifies the optimal control action for each state $(x_{N - 1}, \tau_{N - 1})$, yielding both $\pi_{N - 1}^*$ and $J_{N - 1}$. Continuing this process for all $k \in \{N - 2, \dots, 0 \}$ yields the full sequence $\pi^*$. In practice, performing this backward recursion requires discretizing and bounding the state and action spaces, yielding an approximately optimal policy which we refer to as $\pi^{\rm DP}$. A visualization of $\pi^{\rm DP}$ is shown in Figure~\ref{fig:policy-plots}.

\subsection{Simple Policies}

While $\pi^{\rm DP}$ is highly performant, its structure may be unintuitive. Moreover, the backward recursion requires the full heat-load distribution and a discretized grid over the joint state and action spaces, whose size grows exponentially with the number of compressors; it therefore quickly becomes intractable for larger facilities. Hence, in the following sections, we present two policies that require little information about the heat-load distribution and can be easily computed as alternatives to $\pi^{\rm DP}$. 
Motivated by the specific facility considered in the case study (Section~\ref{sec:case_study}), we present the policies in a two-compressor setting where $\left\lvert \mathcal{C} \right\rvert = 2$, the mean demand exceeds the full-load capacity of compressor \(c_1\), and the maximum demand remains within the combined compressor capacity:\footnote{Although written in terms of two compressors, the same policy structure can be applied to larger systems by treating $c_1$ as an aggregate of multiple base compressors operated at full capacity and $c_2$ as the trim compressor.}
\begin{equation}
    q_{c_1}^{\max} < \bar{q}_{\mathrm{in}} \triangleq \mathbb{E} \left[ Q_{\mathrm{in}} \right]
    \leq q_{\mathrm{in}}^{\max}
    \leq q_{c_1}^{\max}+q_{c_2}^{\max}.
    \label{eq:simple-policy-regime}
\end{equation}
Here, we assume that $\mathcal{C} = \{ c_1, c_2 \}$ is ordered by full-load efficiency, i.e.,
\begin{equation}
\frac{W_{c_1}(q_{c_1}^{\max})}{q_{c_1}^{\max}} \leq \frac{W_{c_2}(q_{c_2}^{\max})}{q_{c_2}^{\max}}.
\end{equation}
For both policies, we operate the more efficient compressor $c_1$ at maximum capacity
\begin{equation}
\pi^{\mathrm{FL}}_{c_1,k}(x_k,\tau_k)
=
\pi^{\mathrm{PL}}_{c_1,k}(x_k,\tau_k)
=q_{c_1}^{\max}
\end{equation}
and focus only on how to operate the second, less efficient compressor $c_2$. This structure is motivated by
\begin{enumerate}[label=(\roman*)]
    \item the fact that each compressor operates most efficiently at its maximum capacity, and
    \item the result from \cite{konda2024} demonstrating that, when ordered by full-load efficiency, a static compressor sequence can achieve minimum cost in the lossless setting.
\end{enumerate}

\vspace{0.2cm}
\noindent \textbf{Full-Load Policy (FL):} The full-load policy, denoted by $\pi^{\mathrm{FL}}$, toggles the second compressor between two operational modes: fully on or fully off. Its specific form is given by
\begin{align}
\pi^{\mathrm{FL}}_{c_2,k}(x_k,\tau_k) &=
    \begin{cases}
        q_{c_2}^{\max}
        &
        \tau_{c_2,k}>0 \; \text{or }
        x_k < \theta_{\mathrm{off}} \\
        0
        &
        \text{otherwise},
    \end{cases}
\label{eq:staged-threshold-policy}
\end{align}
where
\begin{equation}
\theta_{\mathrm{off}} \triangleq 
\begin{cases}
    \frac{1}{\rho} \left( q_{\mathrm{in}}^{\max}-q_{c_1}^{\max} \right) & \rho \in (0, 1] \\
    \infty & \rho = 0    
\end{cases}
\end{equation}
is the minimum inventory level needed to meet the worst-case disturbance $q_{\mathrm{in}}^{\max}$ without turning on $c_2$.
%
Hence, whenever inventory is insufficient to meet the worst-case demand, the compressor operates at full capacity; otherwise, it remains off. Importantly, the FL policy requires knowledge of only the maximum possible heat disturbance $q_{\mathrm{in}}^{\max}$. Its behavior is visualized in Figure~\ref{fig:policy-plots}.

\vspace{0.2cm}
\noindent \textbf{Partial-Load Policy (PL):} Although most efficient per unit of cooling, a minimum on-time after startup can make full-load operation pre-cool more than is economical in a lossy system. Hence, we consider the partial-load policy, denoted by $\pi^{\mathrm{PL}}$, which allows the second compressor to operate at partial load.

For ease of presentation, we first describe a simpler policy for the special case where $q_{\mathrm{in}, k} = \bar{q}_{\mathrm{in}}$ for all $k \in \{0, \dots, N-1\}$. In this setting, the second compressor is operated to cycle between three operational modes:
\begin{itemize}
    \item Partial load: the compressor supplies $q_{c_2}^{\mathrm{avg}} \triangleq \max\{\bar{q}_{\mathrm{in}} - q_{c_1}^{\max},\, q_{c_2}^{\min}\}$, the mean residual demand raised to the compressor's minimum capacity if necessary. This mode lasts for $H_{c_2} - l^\star$ consecutive periods, where $l^\star \in \{0, \dots, H_{c_2}\}$.
    \item Full load: the compressor supplies $q_{c_2}^{\max}$. This mode lasts for $l^\star$ consecutive periods.
    \item No load: the compressor supplies no cooling. This mode lasts for $n_{\mathrm{off}}(l^\star)$ consecutive periods.
\end{itemize}
Here, $n_{\mathrm{off}}(l^\star)$ is the largest number of off periods that can follow the first two modes without making inventory negative, which can be computed in closed form through iterates of the inventory dynamics \eqref{eq:v3-inventory-dynamics}. The optimal time split $l^\star$ is then chosen to minimize the average electrical work per stage:
\begin{equation}
l^\star\in
\argmin_{l \in \{0,\ldots,H_{c_2}\}}
\frac{(H_{c_2} - l) W_{c_2}(q_{c_2}^{\mathrm{avg}})
+ l W_{c_2}(q_{c_2}^{\max})}{H_{c_2} + n_{\mathrm{off}}(l)}.
\label{eq:online-time-split}
\end{equation}

This time split $l^\star$ is what defines the true policy, which we now describe. Actual demand varies, so instead of supplying a fixed $q_{c_2}^{\mathrm{avg}}$ when operating in partial load, compressor $c_2$ provides
\begin{equation}
    q_{c_2}^{\mathrm{par}}(x_k) = \max\{ q_{c_2}^{\mathrm{avg}}, q_{\mathrm{in}}^{\max}-q_{c_1}^{\max}-\rho x_k \}
\end{equation}
kWh of cooling so as to ensure the maximum demand $q_{\mathrm{in}}^{\max}$ can be met. The resulting policy is defined as

\begin{equation}
\begin{aligned}
\pi^{\mathrm{PL}}_{c_2,k}(x_k,\tau_{c_2,k}) = 
\begin{cases}
q_{c_2}^{\mathrm{par}}(x_k),
&
\begin{aligned}
&(\tau_{c_2,k}>l^\star), \, \text{or} \\
&(\tau_{c_2,k}=-1,\;
 x_k<\theta_{\mathrm{off}}),
\end{aligned}
\\[12pt]
q_{c_2}^{\max},
&
\begin{aligned}
&(0<\tau_{c_2,k}\le l^\star), \, \text{or} \\
&(\tau_{c_2,k}=0,\;
 x_k<\theta_{\mathrm{off}}),
\end{aligned}
\\[12pt]
0,
&
\tau_{c_2,k}\le0,\quad
x_k\ge\theta_{\mathrm{off}}.
\end{cases}
\end{aligned}
\end{equation}
The PL policy is visualized in Figure~\ref{fig:policy-plots}.

\section{Model Fitting}
\label{sec:model_fitting}

Solving for and evaluating any of the policies described in the previous section requires estimates of the heat-load disturbance $Q_{\mathrm{in}}$, the compressor operating ranges $[q_c^{\min},q_c^{\max}]$, and the compressor work functions $W_c$. We estimate these quantities using one month of telemetry from a real refrigeration site, where we gather
\begin{itemize}
    \item time-varying room temperature measurements and on/off status for each evaporator,
    \item time-varying suction pressure and motor current measurements for each compressor, and
    \item static equipment parameters (evaporator cooling capacities, compressor line voltages, and power factors).
\end{itemize}
Below, we describe how these measurements are used to build models of $Q_{\mathrm{in}}$ and $W_c$.

\subsection{Heat-Load Estimation}

Incoming heat is not directly measured at the facility. Hence, we use the heat absorbed by the evaporators as an indirect proxy for incoming heat:
\begin{equation}
q_{\mathrm{in},k}
=
\sum_{r \in \mathcal{R}}
\sum_{e \in \mathcal{E}_r}
\alpha_e u_{e,k}^{\mathrm{cool}}
\left(
T_{e,k}^{\mathrm{room}}-T_k^{\mathrm{suc}}
\right).
\label{eq:evap-model}
\end{equation}
Here, $u_{e,k}^{\mathrm{cool}}\in\{0,1\}$ indicates whether evaporator $e$ is cooling and $\alpha_e$ is its rated cooling capacity per degree of temperature difference. The suction temperature $T_k^{\mathrm{suc}}$ is directly computed from suction pressure measurements.

We construct $Q_{\mathrm{in}}$ by partitioning all of the observations $\{q_{\mathrm{in},k}\}$ into twenty equal-width bins, assigning each bin's upper edge as its representative heat load and its empirical frequency as the corresponding probability.

\subsection{Compressor Work--Heat Models}

We next estimate the cooling range and electrical work function of each compressor $c\in\mathcal{C}$. Electrical power is estimated from measured motor current according to
\begin{equation}
P_{c,k}^{\mathrm{elec}}
\approx
\frac{\sqrt{3}\,V I_{c,k}\mathrm{PF}}{1000},
\label{eq:power-estimate}
\end{equation}
where $V$ is the line voltage, $I_{c,k}$ is the measured current, and $\mathrm{PF}$ is the power factor. One-minute average electrical power is converted to work $w_{c,k}$ in kWh. We apply an exponential moving average to reduce measurement noise and exclude observations near compressor startup and shutdown.

Treating the heat removed from the evaporators as the incoming heat load as in \eqref{eq:evap-model} implies that thermal inventory is always $0$, so we can estimate the aggregate heat removed by the compressors as $\sum_{c \in \mathcal{C}} q_{c, k} = q_{\mathrm{in}, k}.$ Let $\mathcal{K}_c$ denote the set of times in which compressor $c \in \mathcal{C}$ is the only active compressor. In these periods, we can estimate the heat removed by compressor $c$ as $\widehat{q}_{c, k} = q_{\mathrm{in}, k}$. If $\lvert \mathcal{K}_c \rvert$ is sufficiently large, then we estimate the work-heat curve $W_{c}(q)=a_{c}+b_{c} q$ directly by ordinary least squares:
\begin{equation}
    (a_{c},b_{c})
    \in
    \argmin_{a,b}
    \sum_{k \in \mathcal{K}_c}
    \left(
    w_{c,k}-a-b \widehat{q}_{c, k}
    \right)^2.
    \label{eq:work-heat-fit}
\end{equation}
Otherwise, if compressor $c$ does not have sufficient independent operation data (i.e., if $\lvert \mathcal{K}_c \rvert$ is small), then we use previously identified work functions to estimate the heat removed by compressor $c$ as 
\begin{equation}
    \widehat{q}_{c, k} = q_{\mathrm{in}, k} - \sum_{d \in \mathcal{A}_k \setminus \{c\}} W_d^{-1}(w_{d, k})
\end{equation}
where $\mathcal{A}_k$ is the set of active compressors at time $k$. This yields sufficient data to perform a least-squares estimate of $W_{c}$ as in \eqref{eq:work-heat-fit}.\footnote{In practice, compressors are typically turned on one-by-one in a predefined order (see \cite{konda2024}), so each $W_{c}$ can be sequentially derived in this manner.}

\begin{figure}
    \includegraphics[width=\columnwidth]{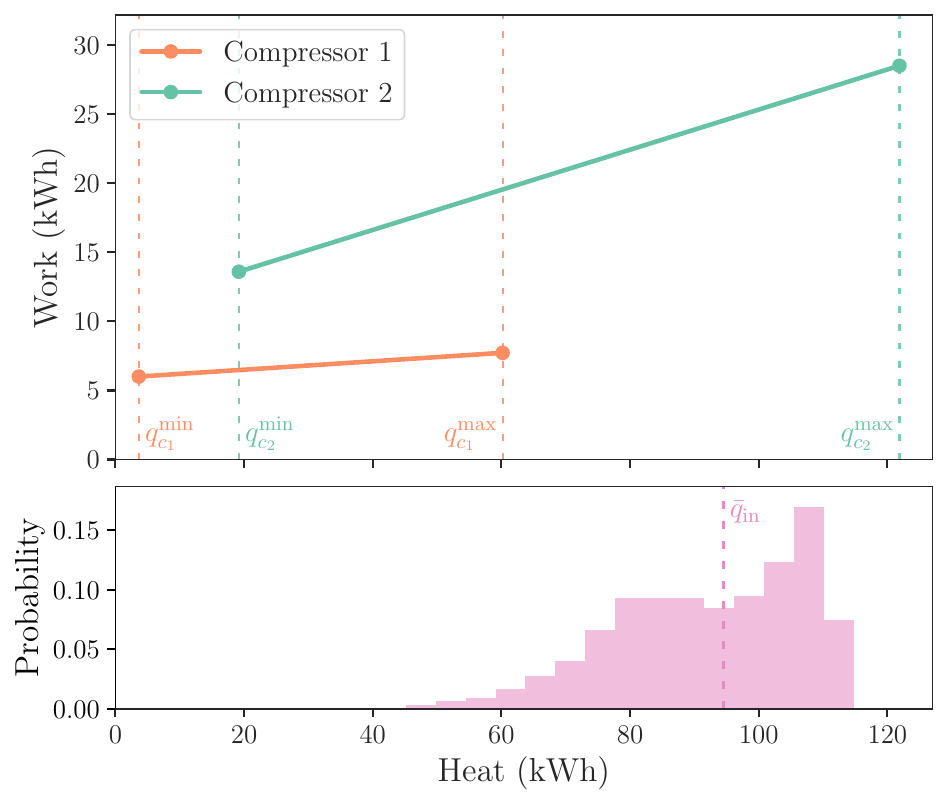}
    \caption{Visualization of fitted model parameters for the Everett facility.
Top: fitted compressor work--heat models over the active operating
ranges $[q_c^{\min},q_c^{\max}]$.
Bottom: empirical distribution of the reconstructed heat-load
observations, with mean $\bar q_{\mathrm{in}}$.}
    \label{fig:everett-data-fit}
\end{figure}

Last, the minimum and maximum cooling capacities $q_{c}^{\min}$ and $q_{c}^{\max}$ of each compressor are taken directly from the data and the curves $W_c$ are restricted to this domain. Figure~\ref{fig:everett-data-fit} summarizes the fitted disturbance and
compressor models for the Everett facility.

\section{Case Study}
\label{sec:case_study}

In this section, we apply and compare the performance of the policies described in Section \ref{sec:algorithms}. Following the approach described in Section \ref{sec:model_fitting}, we estimate the heat load distribution $Q_{\rm in}$ for a cold storage facility in the United States. The facility is equipped with two compressors $\mathcal{C} = \{c_1, c_2\}$ whose work--heat curves we estimate as $W_{c_1}$ and $W_{c_2}$, respectively. The less efficient compressor has a minimum on-time of $H_{c_2}=60$ minutes. The distribution of $Q_{\rm in}$, along with the curves $W_{c_1}$ and $W_{c_2}$, is illustrated in Figure~\ref{fig:everett-data-fit}.

\begin{figure}[h]
    \includegraphics[width=\columnwidth]{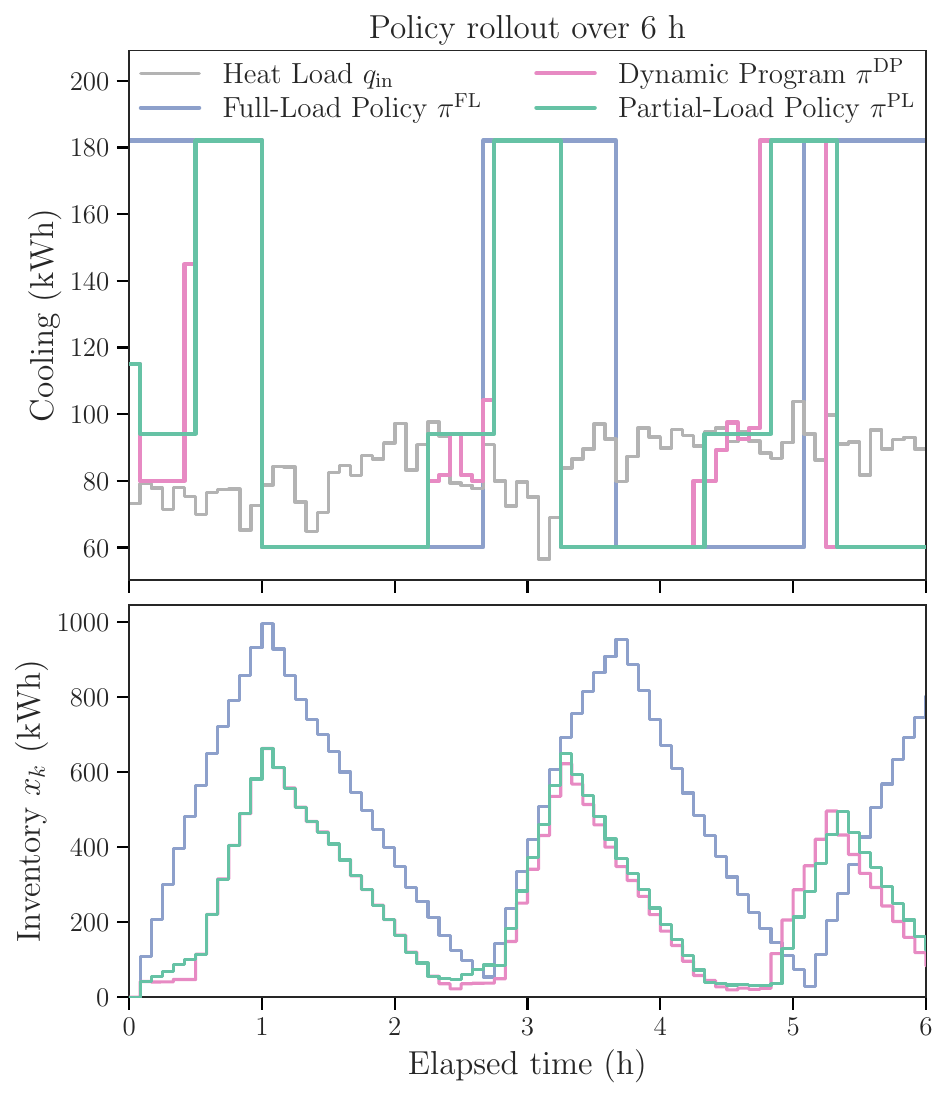}
    \caption{A comparison of the dynamic programming policy $\pi^{\rm DP}$, the full-load policy $\pi^{\mathrm{FL}}$, and the partial-load policy $\pi^{\mathrm{PL}}$. The top plot shows the original heat trajectory (grey) along with the cooling provided by each policy in each period. The bottom plot shows the thermal inventory under each policy.}
    \label{fig:policy-rollout}
\end{figure}

Using these fitted models, we solve \eqref{eq:dp-eq-1}--\eqref{eq:dp-eq-3} to obtain $\pi^{\rm DP}$, use \eqref{eq:staged-threshold-policy} to construct $\pi^{\mathrm{FL}}$, and solve \eqref{eq:online-time-split} to construct $\pi^{\mathrm{PL}}$. All three policies operate the more efficient compressor $c_1$ at full capacity and turn on the less efficient compressor $c_2$ whenever the thermal inventory falls below $\theta_{\mathrm{off}}$. Their operation of $c_2$ during this minimum on-time differs, as shown in Figure~\ref{fig:policy-plots}. For $\rho=0.99$, 1\% of the thermal inventory dissipates each minute. The dynamic program $\pi^{\rm DP}$ therefore initially operates $c_2$ at partial load, maintaining inventory near its minimum feasible level rather than accumulating stored cooling that is unlikely to survive until needed. The partial-load policy $\pi^{\mathrm{PL}}$ mirrors this structure, whereas $\pi^{\mathrm{FL}}$ operates $c_2$ at full capacity throughout the minimum on-time. The six-hour rollout in Figure~\ref{fig:policy-rollout} reflects these differences: $\pi^{\rm DP}$ and $\pi^{\mathrm{PL}}$ both start each minimum on-time at partial capacity and consequently accumulate less inventory than $\pi^{\mathrm{FL}}$.

We further evaluate the policies using 10,000 Monte Carlo simulations with 24-hour heat trajectories sampled from $Q_{\rm in}$. As shown in Table~\ref{tab:monte-carlo-cost}, energy use increases across all policies as $\rho$ decreases and inventory losses grow. For $\pi^{\rm DP}$, reducing $\rho$ from 1 to 0.99 increases the mean cost by 27.3\%, whereas a further reduction to 0.95 adds only 16.7\%, a diminishing marginal increase. Inventory loss also makes continuous full-load operation increasingly costly. 
When $\rho=1$, the partial-load and full-load policies coincide: thermal inventory carries over without loss, so there is no penalty for excess pre-cooling and it is advantageous to exploit the compressor's greater efficiency at full capacity. 
As $\rho$ decreases, however, the cost of carrying excess inventory grows, making partial-load operation increasingly attractive despite its lower instantaneous efficiency. 
Accordingly, $\pi^{\mathrm{FL}}$ costs $2.5\%$ more than $\pi^{\mathrm{PL}}$ at $\rho=0.99$ and $22.7\%$ more at $\rho=0.95$.
By moderating pre-cooling based on the system's lossiness and average demand, $\pi^{\mathrm{PL}}$ remains close to $\pi^{\mathrm{DP}}$, with mean costs only $1.6\%$, $1.0\%$, and $1.8\%$ higher at $\rho=1$, $0.99$, and $0.95$, respectively. In comparison, the mean costs of $\pi^{\mathrm{FL}}$ are 1.6\%, 3.5\%, and 25.0\% higher than those of $\pi^{\rm DP}$ at the same retention factors. At \(\rho=0\), no inventory carries over to the next stage, so pre-cooling provides no benefit. Without knowledge of the next heat load, all three policies must supply at least \(q_{\mathrm{in}}^{\max}\) each period. Both compressors therefore remain on continuously: \(\pi^{\mathrm{DP}}\) and \(\pi^{\mathrm{PL}}\) operate \(c_2\) at \(q_{\mathrm{in}}^{\max}-q_{c_1}^{\max}\), yielding identical costs, whereas \(\pi^{\mathrm{FL}}\) operates it at full capacity and consumes 36.8\% more energy.

\begin{table}[h]
    \centering
    \caption{Mean 24-h cost from Monte Carlo}
    \label{tab:monte-carlo-cost}
    \footnotesize
    \renewcommand{\arraystretch}{1.15}
    \begin{tabular*}{\columnwidth}{@{\extracolsep{\fill}}lcccc@{}}
        \toprule
        & \multicolumn{4}{c}{Mean 24-h cost from Monte Carlo (kWh)} \\
        \cmidrule(lr){2-5}
        Policy
        & $\rho=1$
        & $\rho=0.99$
        & $\rho=0.95$
        & $\rho=0$ \\
        \midrule
        Dynamic Program $\pi^{\rm DP}$
        & 4560.48
        & 5803.71
        & 6771.85
        & 7634.05 \\
        Partial-Load Policy $\pi^{\rm PL}$
        & 4632.55
        & 5860.26
        & 6896.90
        & 7634.05 \\
        Full-Load Policy $\pi^{\rm FL}$
        & 4632.55
        & 6005.58
        & 8462.72
        & 10441.20 \\
        \bottomrule
    \end{tabular*}
\end{table}

To quantify the cost of demand uncertainty, we solve \eqref{eq:dp-eq-1}--\eqref{eq:dp-eq-3} using a 24-hour trajectory of heat loads estimated from real facility data using \eqref{eq:evap-model}. The resulting solution, which we refer to as the clairvoyant DP, is a benchmark that has perfect knowledge of future demand and achieves a cost of 5434.56~kWh for $\rho=0.99$. Although demand uncertainty prevents $\pi^{\rm DP}$ from perfectly anticipating how much thermal inventory will remain useful, it consumes only 3.9\% more energy than clairvoyant DP. Practical implementations, however, generally favor policies that require less computation and less detailed demand information than solving the full dynamic program. The two simple policies demonstrate that this simplicity can be achieved with a modest loss in performance: $\pi^{\mathrm{FL}}$ consumes 6.1\% more energy than $\pi^{\rm DP}$, while $\pi^{\mathrm{PL}}$ consumes only 1.3\% more. Table~\ref{tab:rollout-cost} also shows that the cost of demand uncertainty increases as $\rho$ decreases: $\pi^{\rm DP}$ consumes 1.3\%, 3.9\%, and 6.3\% more energy than clairvoyant DP at $\rho=1$, $0.99$, and $0.95$, respectively. Lower retention shortens the useful lifetime of pre-cooling, making errors in its timing and quantity more costly.

\begin{table}[h]
    \centering
    \caption{24-h rollout cost}
    \label{tab:rollout-cost}
    \footnotesize
    \renewcommand{\arraystretch}{1.15}
    \begin{tabular*}{\columnwidth}{@{\extracolsep{\fill}}lcccc@{}}
        \toprule
        & \multicolumn{4}{c}{24-h rollout cost (kWh)} \\
        \cmidrule(lr){2-5}
        Policy
        & $\rho=1$
        & $\rho=0.99$
        & $\rho=0.95$
        & $\rho=0$ \\
        \midrule
        Clairvoyant DP
        & 4334.46
        & 5434.56
        & 6236.22
        & 6678.34\\
        Dynamic Program $\pi^{\rm DP}$
        & 4391.13
        & 5646.59
        & 6631.76
        & 7634.05 \\
        Partial-Load Policy $\pi^{\rm PL}$
        & 4621.76
        & 5719.58
        & 6776.49
        & 7634.05 \\
        Full-Load Policy $\pi^{\rm FL}$
        & 4621.76
        & 5991.04
        & 8387.28
        & 10441.20 \\
        \bottomrule
    \end{tabular*}
\end{table}

The clairvoyant DP at \(\rho=0\) also approximates the lowest energy consumption achievable without pre-cooling. Relative to this benchmark, \(\pi^{\mathrm{DP}}\), \(\pi^{\mathrm{PL}}\), and \(\pi^{\mathrm{FL}}\) consume 15.4\%, 14.4\%, and 10.3\% less energy at \(\rho=0.99\), demonstrating the benefit of pre-cooling even when 1\% of inventory dissipates each minute. At \(\rho=0.95\), 5\% of inventory dissipates each minute, leaving only 4.6\% after one hour. Consequently, \(\pi^{\mathrm{DP}}\) and \(\pi^{\mathrm{PL}}\) perform close to the benchmark without pre-cooling, while \(\pi^{\mathrm{FL}}\) incurs substantial excess energy consumption. Under such severe losses, there is little benefit to pre-cooling because little stored cooling survives until needed, highlighting the importance of accounting for inventory retention when designing load-shifting strategies.

\section{Conclusion}

This work studies how the lossiness of a refrigeration system affects the costs and structures of compressor control strategies. We model the dynamics of the extra stored cooling in a refrigeration system, where a scalar inventory retention factor determines how much of the stored cooling carries over in each time step. These dynamics lead to an optimization problem where one seeks to minimize the total work of the compressors such that the inventory remains nonnegative while satisfying operational hold time constraints. We present a dynamic program and two simple policies as solutions to this problem, and we analyze their performance on models fit from a real refrigeration facility's data. Our detailed case study shows that when the inventory retention factor is close to one, pre-cooling is an effective strategy, but as it decreases, partial-load operation can offer substantial cost savings; as such, control strategies should be designed with the inventory retention factor in mind.

These results raise a number of important practical issues. We consider a single thermal inventory in this work without discussing the role of the evaporators or how heat and cooling are disaggregated across the rooms. Understanding how to operate the evaporators in tandem with the compressors is crucial for practical pre-cooling at the individual room level. 
Further, the simple policies in this work are developed under the operating regime in~\eqref{eq:simple-policy-regime}, which relates the compressor capacities to the mean and maximum heat loads. Extending these policies beyond this regime and formally characterizing their performance relative to the optimal policy are important directions for future work.
Last, we demonstrate the importance of the inventory retention factor by analyzing the costs of each policy in a variety of settings, but we do not propose a method for estimating the inventory retention factor itself. We believe data can be used in a similar manner to do so, but we leave this estimation problem for future work.

\bibliographystyle{IEEEtran}
\bibliography{ref}

\end{document}